# Large-scale mapping of moiré superlattices by Raman imaging of interlayer breathing mode and moiré phonons

Kai-Qiang Lin*,[†,#], Johannes Holler[†,#], Jonas M. Bauer[†], Marten Scheuck[†], Bo Peng[‡], Tobias Korn[§], Sebastian Bange[†], John M. Lupton[†], Christian Schüller*,[†]

[†]Department of Physics, University of Regensburg, Regensburg, Germany.

[‡]TCM Group, Cavendish Laboratory, University of Cambridge, Cambridge, United Kingdom.

[§] Institute of Physics, University of Rostock, Germany.

KEYWORDS: *moiré superlattice, interlayer breathing mode, moiré phonon, low-frequency Raman scattering, hyperspectral Raman imaging.*

ABSTRACT

Moiré superlattices can induce correlated-electronic phases in twisted van-der-Waals materials – strongly correlated quantum phenomena emerge, such as superconductivity and the Mott-insulating state. However, moiré superlattices produced through artificial stacking can be quite inhomogeneous, which hampers the development of a clear correlation between the moiré period and the emerging electrical and optical properties. Here we demonstrate in twisted-bilayer transition-metal dichalcogenides that low-frequency Raman scattering can be utilized not only to detect atomic reconstruction, but also to map out the inhomogeneity of the moiré lattice over large areas. The method is established based on the finding that both the interlayer-breathing mode and moiré phonons are highly susceptible to the moiré period and provide characteristic fingerprints. We visualize microscopic domains with an effective twist-angle resolution of ~0.1°. This ambient non-invasive methodology can be conveniently implemented to characterize and preselect high-quality areas of samples for subsequent device fabrication, and for transport and optical experiments.

Long-range periodicity arising from the moiré potential landscape has enabled fundamental changes to the electronic and phononic properties of van der Waals homo- and heterostructures[1-8]. The period of the moiré superlattice can be conveniently tuned by twist angle, but is, in general, quite challenging to characterize. Considerable effort has been invested into mapping of the moiré superlattice. Transmission electron microscopy (TEM) can visualize moiré superlattices at the ultimate spatial resolution[9, 10]. Unfortunately, the high-energy electron beam tends to create defects in these two-dimensional materials, and TEM necessitates specific sample preparation, such as suspension or support by thin membranes, which introduces strain and is not fully compatible with subsequent device fabrication and transport measurements. Various scanning probe microscopy (SPM) methods have been developed to visualize moiré superlattices[11-14]. In these measurements, the top surface of the twisted bilayers must be exposed to the tip, while most high-quality twisted bilayer samples are still fabricated with top and bottom hBN encapsulation for transport measurements and optical spectroscopy. A convenient method that can map out the moiré period over large areas and is also applicable to encapsulated layers is thus still lacking. Low-frequency Raman spectroscopy has emerged as a powerful tool to characterize van der Waals homo- and heterostructures through their interlayer breathing mode and shear modes[15, 16], which offer rich information such as the number of layers and the stacking configuration. In twisted structures, though, the atomic periodicity of the two adjacent layers no longer matches in an in-plane direction[17]. The shear mode is thus generally not expected to contribute to Raman spectra of twisted bilayers[18]. The interlayer breathing mode, on the other hand, could conceivably persist even in twisted structures[19, 20]. Recent calculations[6] have suggested that the interlayer breathing mode can be sensitive to the twist angle and therefore potentially enable its determination. It remains somewhat unclear, however, how a moiré superlattice as sketched in the schematic of Figure 1a influences the interlayer breathing mode in an experiment.

Here, we show that low-frequency Raman scattering of the interlayer breathing mode and moiré phonons in bilayer transition metal dichalcogenides (TMDCs) offers a uniquely sensitive probe of the twist angle. We demonstrate a significant shift of the frequency of the breathing mode with changes to the moiré period, which is quantified simultaneously by the frequency of the moiré phonons. The approach provides a convenient method to map out the microscopic inhomogeneity of the moiré superlattice with diffraction-level resolution over large areas. Hyperspectral imaging of a 5° twisted bilayer of $WSe_2$ allows us to distinguish individual domains featuring loose interfacial contact, local rotational sliding motion, and atomic reconstruction over a total sample area exceeding 1,000 μm$^2$.

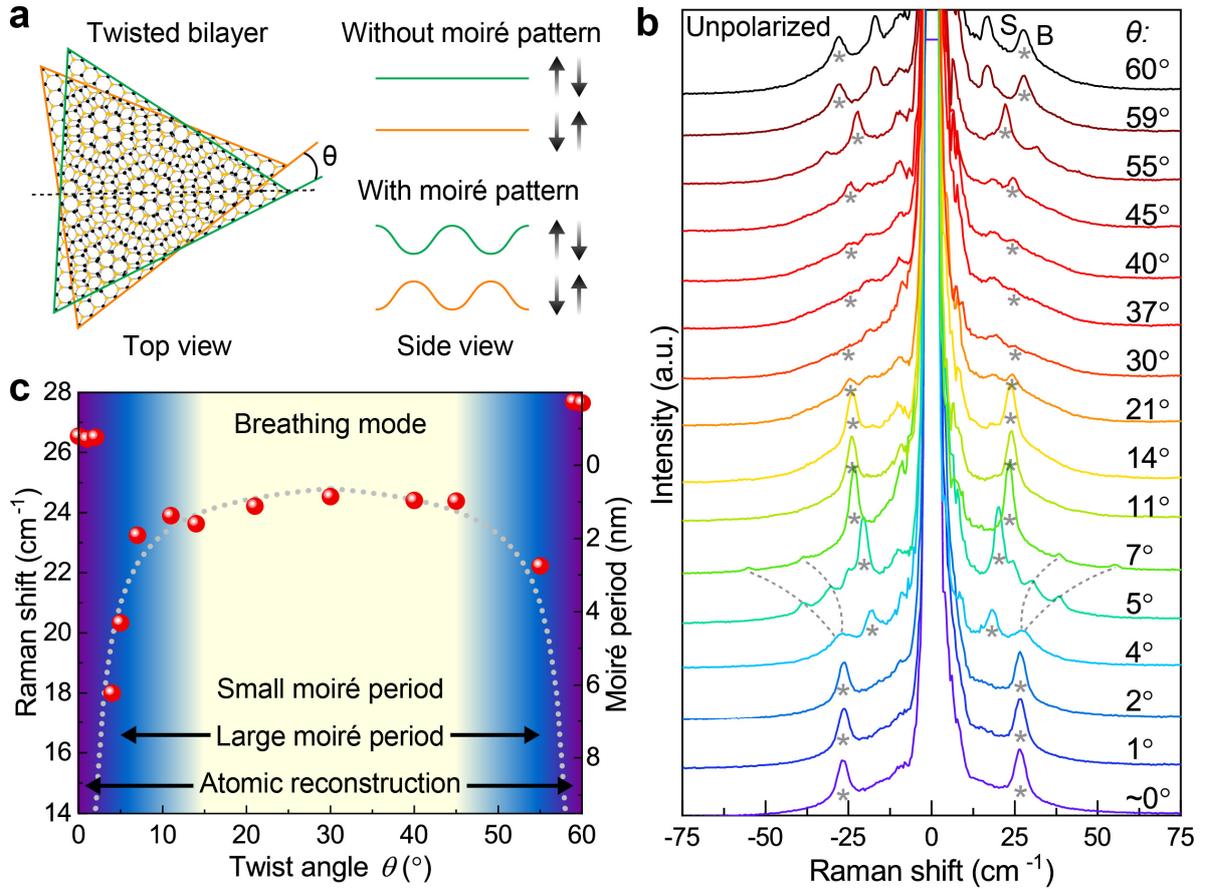

**Figure 1.** Twist-angle dependence of low-frequency Raman scattering. (a) Illustration of twisted bilayer WSe$_2$ in real space and the low-energy interlayer breathing mode with and without the presence of a moiré pattern. The green and orange lines sketch the landscape of the top and bottom layers. (b) Low-frequency Raman spectra of bilayer WSe$_2$ as a function of twist angle. The interlayer breathing modes are marked by asterisks and the dashed lines indicate the moiré phonon. The breathing mode (B) and shear mode (S) from the 2H natural bilayer (60°) are shown for reference. (c) The twist-angle dependence of the breathing-mode frequency (red dots) clearly distinguishes three regimes of twist angles: atomic reconstruction (purple), where the breathing mode stiffens; large moiré unit cells (blue), where the breathing mode softens with increasing moiré period; and small moiré unit cells (yellow tint), where the breathing mode remains near

constant with twist angle. The grey-dotted curve shows the twist-angle dependence of the moiré period. The error from fittings is below the diameter of the spheres.

We begin by fabricating twisted-bilayer WSe$_2$ to investigate how the interlayer breathing mode and moiré phonons change with the twist angle, i.e. the moiré period. As detailed in the Methods Section, the samples are prepared through mechanical exfoliation and a sequential deterministic dry-transfer technique[21], with the twist angle confirmed by polarization-resolved second-harmonic generation (SHG)[22]. Figure 1b shows low-frequency Raman spectra of sixteen twisted-bilayer WSe$_2$ samples with different twist angles between 0° (3R stacking) and 60° (2H stacking). Prominent peaks appear below 50 cm$^{-1}$, where the low-frequency breathing and shear modes are generally expected[15]. The measurement is carried out with an unpolarized detection configuration as detailed in the Methods Section and in Figure S1a. It has been shown in natural multilayers that the breathing mode can only be observed under the condition where excitation and detection are polarized in parallel, while the shear mode can be observed in both parallel- and cross-polarized arrangements[15, 23-25]. By comparing parallel- and cross-polarized Raman scattering spectra in Figure S1b, we identify the breathing mode of twisted bilayers in Figure 1b and mark this by an asterisk.

Although commensurate crystallographic superlattices can only emerge at certain twist angles, incommensurate quasicrystalline moiré patterns can form for nearly every twist angle[4, 26]. The moiré period in homobilayers is defined as $\lambda = 0.5\ a/\sin(\theta/2)$,[26] where $a$ is the in-plane lattice constant and $\theta$ is the effective twist angle between the bilayers. Since TMDC monolayers have a C$_3$ symmetry, $\theta$ is smaller than 30°. We first discuss the bilayers with twist angles that are between 0° and 3°, where, in an ideal case, the moiré period can be expected to be largest. Surprisingly, these samples all show an identical breathing mode at 27 cm$^{-1}$ (marked by asterisks in Figure 1b), independent of the twist angle. The same independence on twist angle is also observed for the interlayer shear mode shown in Figure S1b. This independence can be interpreted as either indicating that the breathing mode is insensitive to the moiré superlattice; or else that a well-defined moiré superlattice does not actually form in these samples. On the other hand, starting from twist angles of 4°, the interlayer breathing mode does show a significant twist-angle dependence. At the same time, additional peaks (marked by dashed lines in Figure 1b) arise at higher frequencies. A continuous blue-shift of the breathing mode with increasing twist angle is observed in bilayers up to twist angles of ~14°, beyond which their intensity drops significantly. For bilayers with twist angles between 14° and 45°, the frequency of the breathing mode remains nearly constant and its intensity stays weak. From 30° onwards, the moiré period is expected to increase again.[4, 26] For the bilayer with 55° twist angle, the intensity of the breathing mode recovers, and the frequency simultaneously shifts to the red, showing a similar Raman spectrum as the 5° twisted bilayer. For the 59° twisted bilayer, the breathing mode suddenly jumps to a significantly higher frequency, beyond that of the ~0° twisted bilayer but identical to the frequency found for 2H natural (60°) bilayers of WSe$_2$. Figure 1c summarizes the twist-angle dependence of the breathing-mode frequencies (red spheres) as extracted from the Raman spectra. Intriguingly, this twist-angle dependence clearly coincides with the calculated moiré period (grey dotted line).

To experimentally quantify the moiré period, we examine the moiré phonons, which were first observed by Lin *et al.* in twisted-bilayer MoS$_2$[4]. Moiré phonons refer to phonon modes that are outside the zone center Γ and Raman inactive in natural bilayers but become active in twisted bilayers due to the folding of the phonon bands by the moiré superlattice. Figure 2a shows the Raman spectra of twisted-bilayer WSe$_2$ over a broad frequency range, with peaks assigned to moiré phonons marked by arrows. These peaks feature significant shifts with the variation of the twist angle. Figure 2b summarizes the correlation between peak positions and twist angles. Following the methodology documented in Ref. 4, we calculate the folded longitudinal acoustic (LA) phonon and transverse acoustic (TA) phonon of twisted-bilayer WSe$_2$ based on the phonon dispersion of the monolayer WSe$_2$ described in Ref. [27]. As shown in Figure 2b, the calculated folded LA phonon (red line) and TA phonon (blue line) branches overlap well with the experimental peak positions. Deviations between the experimental points and the calculation may arise due to the fact that the phonon dispersion is calculated for an unstrained monolayer. Slight variations of the phonon frequencies due to the van der Waals interaction between the layers are therefore possible. Also, in the calculation, the interactions between the layers with the substrate are not taken into account.

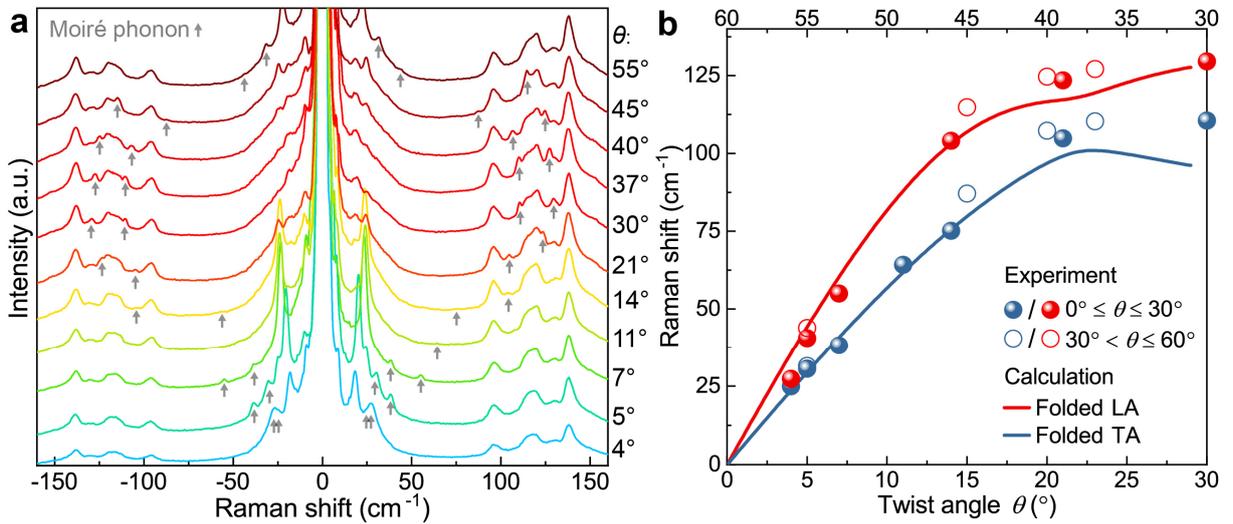

**Figure 2.** Moiré phonons in twisted-bilayer WSe$_2$. (a) Raman spectra of bilayer WSe$_2$ for different twist angles, spanning a broader frequency range than in Figure 1. Peaks assigned to moiré phonons are marked by arrows. (b) Comparison between the experimental moiré-phonon

frequency with the calculated folded LA and TA phonon branches as a function of twist angle. The experimental error is below the diameter of the spheres.

The appearance of moiré phonons in the Raman spectra offers direct evidence for the formation of the moiré superlattice. The concurrent appearance of the moiré-phonon bands and the shifting of the breathing mode with twist angle indicates that the breathing mode is indeed sensitive to the moiré superlattice. Observation of the same interlayer breathing and shear-mode frequencies for twisted bilayers with twist angles smaller than 3° provides straightforward evidence of atomic reconstruction, which was recently observed in stacked van der Waals homo- and heterostructures employing a range of microscopic techniques[6, 10, 18, 28-30]. With reconstruction, the atomic sites relax to a lower-energy configuration and, depending on the twist angle, form domains that follow the 2H- or 3R- stacking geometry. Assuming that such reconstruction is effective for the 59° twisted-bilayer $WSe_2$ sample, the interlayer breathing-mode frequency would indeed be expected to closely resemble that of the natural 2H-stacking bilayer $WSe_2$. The ~3° threshold for this reconstruction appears to agree with recent experimental observations[18, 29, 30] as well as calculations[6, 28]. When the twist angle rises above this threshold, the moiré superlattice becomes stable and leads to a significant softening of the interlayer breathing mode. This mode then stiffens again with decreasing moiré period, i.e. with increasing twist angle. Such a correlation between the interlayer breathing mode and the twist angle is rather unexpected from the perspective of the interlayer spacing alone, since bilayers with a smaller twist angles are expected to have the smaller interlayer distance and therefore a larger interlayer force constant in the picture of a linear chain model[31]. The correlation can be rationalized by considering the mixing between in-plane and out-of-plane modes[6]. When the moiré pattern forms, the moiré superlattice is no longer flat but acquires a periodic corrugation[32] as illustrated in Figure 1a, which enables mixing of the out-of-plane and the in-plane displacement modes. Indeed, in the 4° and 5° twisted bilayers, the intensity of the interlayer breathing modes is not completely suppressed in a cross-polarization configuration, in stark contrast to the case of bilayers with larger twist angles as shown in Figure S1b. For bilayers with twist angles between 14° to 45°, the constant frequency of the interlayer breathing mode coincides with a minor change of the moiré period in this angular range. The interlayer breathing-mode frequency can thus clearly identify three distinct regimes of interlayer coupling, marked in Figure 1c: the occurrence of atomic reconstruction (purple), and the presence of large (blue) and small moiré periods (yellow). Crucially, both the interlayer breathing mode as well as the folded acoustic phonons appear to have the largest susceptibility to twist angle in the regime of large moiré periods, where strong correlation phenomena are expected to emerge.

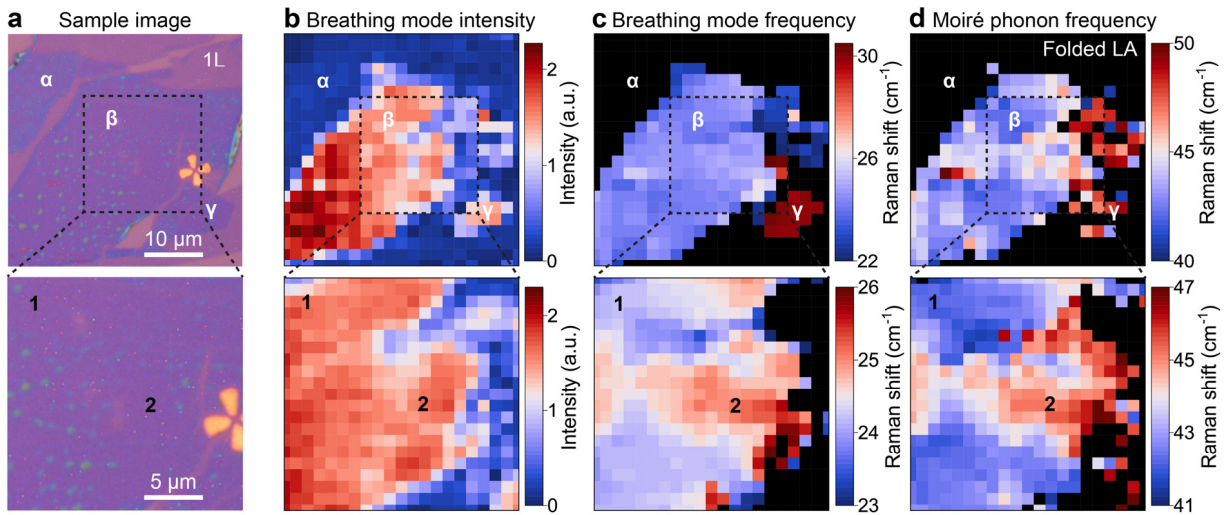

**Figure 3.** Low-frequency Raman mapping of a 5° twist-angle $WSe_2$ bilayer over a 40×40 μm² area. (a) Reflection image of the sample (dark purple) under white-light illumination. The single-layer region (light purple) is marked by "1L". (b) Spatially resolved intensity of the interlayer breathing mode for the same sample area. (c) Spatially resolved Raman frequency of the interlayer

breathing mode. (d) Spatially resolved Raman shift for the folded LA phonon. For each column, the lower panels show a 20×20 μm$^2$ close-up region as marked in the upper panels. The Raman mapping uses a step size of 2 μm in the top panels and 1 μm below. The Raman frequency of the interlayer breathing mode distinguishes three different areas α, β, and γ. Domains 1 and 2 in the β area are distinguished by the Raman frequencies of both the breathing mode and the moiré LA phonon.

Next, we explore the possibility of using the breathing mode and the moiré phonons to map out inhomogeneities of twisted bilayers. Figure 3a shows microscope images of a $\theta$=5° twisted-bilayer WSe$_2$ sample. For WSe$_2$ homobilayers, twist angles between 4° and 5.1° are particularly interesting, since for this range of angles, a series of highly correlated electron phases has been reported[5]. Areas in the optical microscopy image that are dark purple correspond to the twisted bilayer, whereas the light-purple area indicates the monolayer region (marked 1L). We perform hyperspectral Raman imaging over the entire sample area. Figures 3b-c shows the spatially resolved intensity and frequency of the breathing mode, delivering rich information at a spatial resolution of 2 μm (top panels) and 1 μm (bottom panels). Comparing Figure 3a-c, we identify three different areas of the bilayer labeled α, β, and γ. As for the monolayer area, the α area does not show a breathing mode, although this sample region is identified clearly as a bilayer from the optical microscopy images in Figure 3a and Figure S2. We attribute this absence of the interlayer breathing mode to a loose contact between the layers in the α area. Such cleavage can occur when ambient humidity is sufficiently high that a thin water layer may form between layers during the stamping process[33, 34]. In contrast, both β and γ bilayer areas show intense interlayer breathing modes, but the vibrational frequency in these two sample areas differs by more than 8 cm$^{-1}$, which is far beyond the small local variation in twist angle. Instead, the Raman spectrum of the "γ" bilayer region is similar to that of a $\theta \approx 0°$ twisted-bilayer WSe$_2$, with a breathing mode frequency close to 30 cm$^{-1}$ and no resolvable moiré phonons. This absence indicates that significant rotational sliding and atomic reconstruction occur in the γ area. With such rotational sliding in mind, we further characterize the local variations of the twist angle in the close-up region around the β area in the lower panels. As shown in the bottom panel of Figure 3c, distinct domains (marked 1 and 2) show up in this β area and exhibit different and distinct breathing-mode frequencies. The same domains are observed in the spatial distribution of the moiré-phonon frequencies in the bottom panel of Figure 3d. This imaging based on the moiré phonon is significantly noisier than that based on the interlayer breathing mode in panel c, which can be attributed to the order of magnitude reduction in Raman intensity from the moiré phonons compared to the interlayer breathing modes. Therefore, within a limited acquisition time, imaging based on the interlayer breathing mode has the advantage of an improved signal-to-noise ratio. Finally, we plot typical Raman spectra for the areas α, β, and γ in Figure 4a, and for domains 1 and 2 of the β region in Figure 4b. The Raman spectra from the bilayer areas α, β, and γ differ strongly, while the spectra from domains 1 and 2 are rather similar. The interlayer breathing mode and folded LA and TA phonons are all present in the spectra for both domains in the β sample area, although there are clear shifts in their frequencies. The frequency difference of the folded LA phonon between domains 1 and 2 is about 3.5 cm$^{-1}$, which corresponds to a variation in twist angle of ~0.4° around the nominal value of 5° according to the dispersion plotted in Figure 2b. A typical Raman spectrometer with a high-resolution grating can easily reach a resolution of below 0.5 cm$^{-1}$, implying a resolution in twist angle of better than 0.06°.

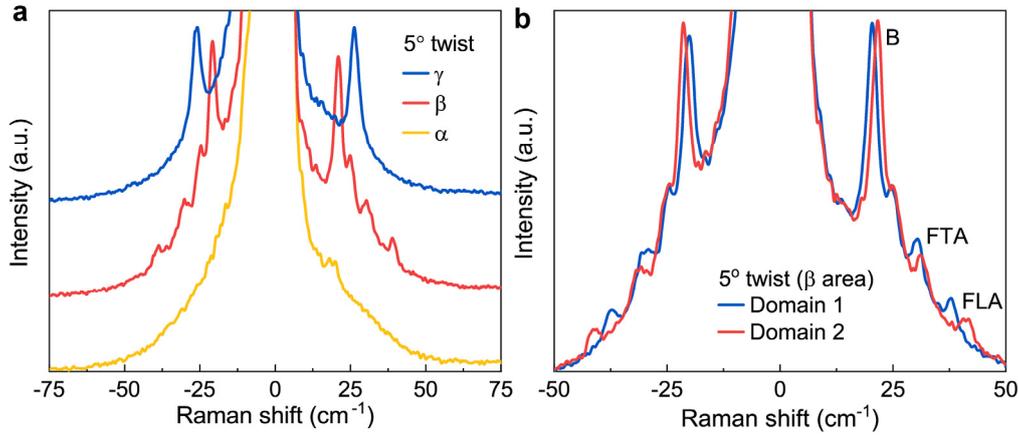

**Figure 4.** Low-frequency Raman spectra of different sample areas and stacking domains of 5° twisted-bilayer WSe$_2$. (a) Representative Raman spectra of areas α, β, and γ marked in Figure 3. (b) Representative Raman spectra of domains 1 and 2 within the β area, showing different Raman frequencies for both the interlayer breathing mode (labelled "B") and the moiré phonons (labelled "FLA" for the folded LA phonon and "FTA" for the folded TA phonon).

In summary, we have demonstrated that low-frequency Raman scattering can probe the local moiré period in twisted bilayer TMDCs through both the interlayer breathing mode and moiré phonons. The moiré superlattice homogeneity can be mapped out conveniently across a sample area in excess of 1,000 μm$^2$. We find that twisted bilayers fabricated by mechanical exfoliation and deterministic stacking can exhibit inhomogeneities over length scales of tens of micrometers. Twist angles in the stacked bilayer can deviate locally from the value inferred from the difference between the crystal orientations of the individual monolayers as determined by polarization-resolved SHG. Hyperspectral Raman imaging can clearly distinguish the different areas, identifying loose interfacial contact, atomic reconstruction, and rotational sliding. Such ambient optical techniques offer a straightforward method of precharacterizing the moiré superlattice on large length scales in order to identify high-quality areas with well-defined twist angles prior to device fabrication. Knowledge of the twist-angle dependence of the interlayer breathing mode can be directly transferred to tip-enhanced Raman spectroscopy to spatially resolve the moiré superlattice on even finer length scales below the supercell level.

**Experimental Methods.** *Sample preparation and twist angle pre-determination.* Monolayer $WSe_2$ and $MoSe_2$ flakes were exfoliated from bulk crystals (HQ Graphene) onto commercial polydimethylsiloxane PDMS films (Gel-Pak, Gel-film X4) using Nitto tape (Nitto Denko, SPV 224P)[21]. To obtain twisted TMDC bilayers, one single monolayer flake was first partially stamped onto a silicon chip with a 285 nm $SiO_2$ layer on top. The remaining part of the monolayer flake was then transferred on top of the first flake after rotating the silicon chip to a desired twist angle $\theta$. The two stamping processes were carried out sequentially using an optical microscope combined with translation stages. The silicon chip was placed on a Peltier heating stage, which was set to 65 °C during the stamping. Figure S2 shows an example of a 5° twisted-bilayer $WSe_2$ sample resulting from this process. The twist angles are further characterized by measuring the crystal orientation of the monolayer areas of the individual subsections. This orientation is determined by measuring the co-polarized SHG intensity as a function of the relative angle between crystal axis and laser polarization[22].

*Raman spectroscopy.* A continuous-wave, 532 nm laser was focused down to a ~1 μm diameter spot on the sample through a 0.9 numerical-aperture microscope objective (Nikon, 100×) under ambient condition. The excitation power was set to 2.5 mW and the Raman scattering signal was collected by the same objective. After passing through a 50:50 beam splitter and a set of Bragg filters, the signal was dispersed by an 1800 grooves/mm grating and detected by a CCD camera.

For cross-polarization measurements, a polarizer was placed right after the Bragg filters. The integration time was set to 60 s for unpolarized Raman-scattering measurements (in Figures 1b and 2a) and 180 s for the cross-polarized Raman-scattering measurement (in Figure S1b). For hyperspectral Raman imaging (Figures 3 and 4), the integration time was set to 10 s per spot.

ASSOCIATED CONTENT

**Supporting Information**.

The Supporting Information is available free of charge.

AUTHOR INFORMATION

**Corresponding Author**

*E-mail: kaiqiang.lin@ur.de; christian.schueller@ur.de;

**Author Contributions**

K.L. and C. S. led the project. K.L. conceived the work. J. H. and K.L. performed measurements. J.M.B. and K.L. prepared the samples. K.L. processed and analyzed the data, and wrote the manuscript with contributions of all authors. All authors discussed results and contributed to the analysis, and have given approval to the final version of the manuscript. #These authors contributed equally.

**Notes**

The authors declare no competing financial interest.

ACKNOWLEDGMENT

This work was supported by the Deutsche Forschungsgemeinschaft (DFG, German Research Foundation) SPP 2244 (Project-ID LI 3725/1-1, 443378379, and SCHU1171/10-1, 443361515), and SFB 1277 (Project-ID 314695032) projects B03, B05 and B06. T.K. gratefully acknowledges support by the DFG via KO 3612/4-1.

TOC image:

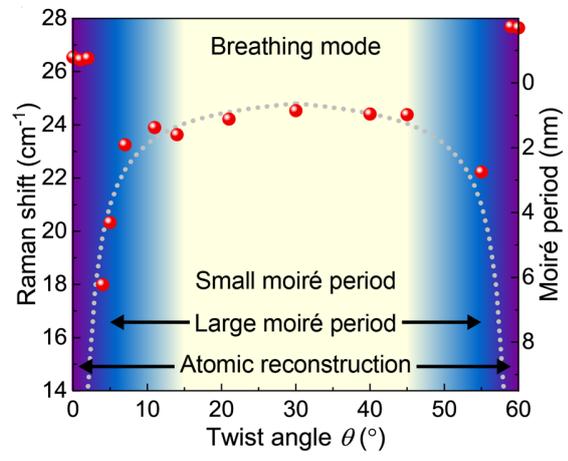